\documentclass[11pt]{article}

\usepackage[final]{acl}

\usepackage{times}
\usepackage{latexsym}

\usepackage[T1]{fontenc}

\usepackage[utf8]{inputenc}

\usepackage{microtype}

\usepackage{inconsolata}

\usepackage{graphicx}

\usepackage{titlesec}

\titlespacing{\subsubsection}{0em}{1.25ex plus 1ex minus .2ex}{0.5em}

\usepackage{enumitem} 
\usepackage{amssymb}
\usepackage{multirow}
 \usepackage{booktabs}
\usepackage{amsmath}
\usepackage{array}
\usepackage{tabularx} 
\usepackage{pgfplots} 
\pgfplotsset{compat=1.18}
\usepgfplotslibrary{statistics}
\usepackage{algorithm}
\usepackage{algorithmic}
\usepackage{xcolor}          

\definecolor{myblue}{RGB}{78, 121, 167}   
\definecolor{myorange}{RGB}{242, 142, 43} 
\definecolor{myred}{RGB}{225, 87, 89}    

\usepackage{listings}
\usepackage[most]{tcolorbox}

\definecolor{backcolour}{rgb}{0.97,0.98,0.99}
\definecolor{keycolor}{rgb}{0.0, 0.3, 0.6}
\definecolor{stringcolor}{rgb}{0.0, 0.5, 0.0}
\definecolor{commentcolor}{rgb}{0.5, 0.5, 0.5}
\definecolor{framecolor}{rgb}{0.7, 0.75, 0.8}

\lstdefinelanguage{yaml}{
  keywords={true,false,null,y,n},
  keywordstyle=\color{darkgray}\bfseries,
  basicstyle=\ttfamily\footnotesize,
  sensitive=false,
  comment=[l]{\#},
  morecomment=[s]{/*}{*/},
  commentstyle=\color{commentcolor}\itshape,
  stringstyle=\color{stringcolor},
  moredelim=[l][\color{orange}]{\&},
  moredelim=[l][\color{magenta}]{*},
  moredelim=**[il][\color{keycolor}]{:},
  morestring=[b]',
  morestring=[b]",
    keepspaces=true,         
  columns=fixed,           
  basewidth=0.5em,         
  showstringspaces=false,  
  showspaces=false,        
  tabsize=2,               
  breaklines=true          
}

\newtcolorbox{contractbox}[2][]{
    enhanced,
    width=\linewidth,     
    colback=backcolour,
    colframe=framecolor,
    boxrule=0.8pt,
    arc=3mm,
    fonttitle=\bfseries\small\sffamily,
    coltitle=black,
    attach boxed title to top left={xshift=4mm, yshift=-3mm},
    boxed title style={
        colback=white,
        colframe=framecolor,
        boxrule=0.8pt,
        arc=2mm
    },
    title={#2},
    #1
}

\definecolor{cvprblue}{RGB}{33, 33, 33}       
\definecolor{promptbg}{RGB}{245, 245, 245}    
\definecolor{framegray}{RGB}{128, 128, 128}   

\newtcolorbox{promptbox}[2][]{
  enhanced,
  title={#2},                     
  colback=promptbg,               
  colbacktitle=framegray,          
  coltitle=white,                 
  colframe=framegray,              
  boxrule=0.5pt,                  
  arc=2pt,                        
  outer arc=2pt,
  fonttitle=\bfseries\sffamily\small, 
  fontupper=\ttfamily\footnotesize\raggedright,    
  left=5pt, right=5pt, top=5pt, bottom=5pt, 
  toptitle=2pt, bottomtitle=2pt,  
  sharp corners=south,            
  #1                              
}

\title{Contract-Coding: Towards Repo-Level Generation via Structured Symbolic Paradigm}

\author{Yi Lin\quad 
        Lujin Zhao\quad 
        Yijie Shi\thanks{\ \ Corresponding author.} \\
  State Key Laboratory of Networking and Switching Technology \\
  Beijing University of Posts and Telecommunications \\
  Beijing, 100876, China \\
  \texttt{\{yilin, zlujin, yijieshi2000\}@bupt.edu.cn} \\}

\begin{document}
\maketitle
\begin{abstract}
The shift toward intent-driven software engineering (often termed "Vibe Coding") exposes a critical Context-Fidelity Trade-off: vague user intents overwhelm linear reasoning chains, leading to architectural collapse in complex repo-level generation. We propose Contract-Coding, a structured symbolic paradigm that bridges unstructured intent and executable code via Autonomous Symbolic Grounding. By projecting ambiguous intents into a formal Language Contract, our framework serves as a Single Source of Truth (SSOT) that enforces topological independence, effectively isolating inter-module implementation details, decreasing topological execution depth and unlocking Architectural Parallelism. Empirically, while state-of-the-art agents suffer from different hallucinations on the Greenfield-5 benchmark, Contract-Coding achieves 47\% functional success while maintaining near-perfect structural integrity. Our work marks a critical step towards repository-scale autonomous engineering: transitioning from strict "specification-following" to robust, intent-driven architecture synthesis. Our code is available at \url{https://github.com/imliinyi/Contract-Coding}.
\end{abstract}

\section{Introduction}

The advent of Large Language Models (LLMs) has fundamentally transformed Automated Software Engineering (ASE)~\cite{Codex, AlphaCode}, evolving from solitary code completion to collaborative Multi-Agent Systems (MAS) capable of handling complex workflows~\cite{AgentVerse, li2023camel}. While frameworks like MetaGPT~\cite{MetaGPT} and ChatDev~\cite{ChatDev} demonstrate that specialized roles enhance problem-solving~\cite{SWE-agent, OpenHands}, they encounter a scalability wall.

A critical gap remains between academic research and real-world adoption. While \textit{Specification-Driven} approaches (Spec-Coding) offer architectural rigor, they suffer from a "Formalization Bottleneck": assuming complete structural blueprints that users rarely possess~\cite{Parsel, CodePlan}. This contradicts the reality of \textit{Intent-Driven} development ("Vibe Coding"), where users provide ambiguous functional goals. Conversely, synthesizing repositories directly from these intents hits a \textit{Context-Fidelity Trade-off}. The dominant "linear chain-of-thought" paradigm~\cite{CoT} faces two systemic bottlenecks: (1) \textit{Sequential Entropy Accumulation}, where early architectural ambiguities amplify as semantic noise~\cite{Reflextion}; and (2) \textit{The Context Bottleneck}, where accumulating implementation details exhausts finite windows, leading to \textit{Architectural Collapse}~\cite{LostinMiddle, MemGPT}.

We challenge the assumption that repository synthesis must be sequential. We argue that the root cause of scalability failure is the entanglement of control complexity and implementation volume. To scale Vibe Coding, we must revisit the principle of \textit{Information Hiding}~\cite{parnas72} via a structured-symbolic lens. Instead of parsing the noisy, full-context raw code of peers, agents should align with a high-fidelity "architectural thumbnail"—or \textit{Language Contract}. This mechanism grounds vague intents into rigorous abstractions~\cite{DesignbyContract}, effectively performing \textit{Semantic Compression} on the reasoning search space to mitigate hallucination risks.

\begin{figure*}[htbp]
\centering
\includegraphics[width=\linewidth]{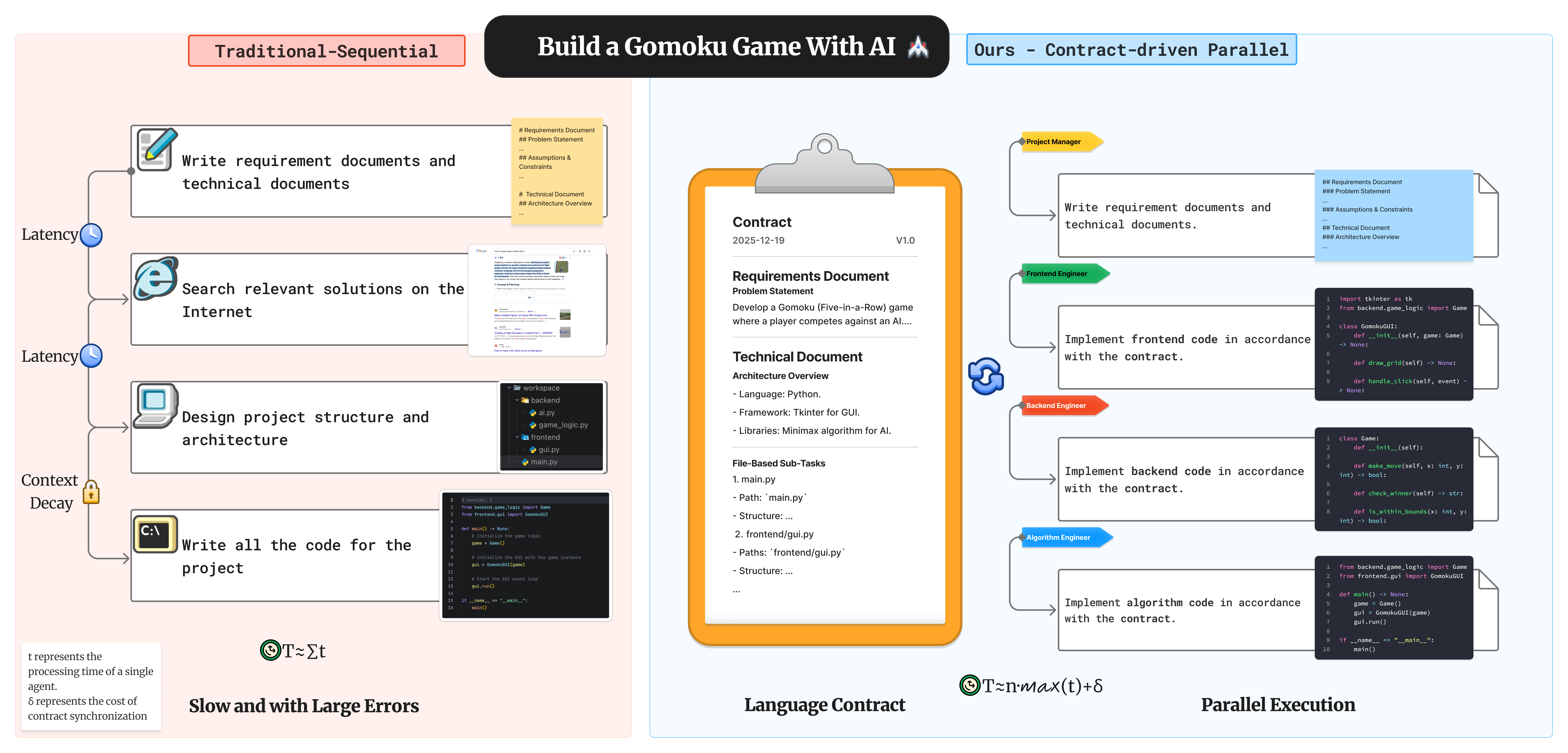}
\caption{Linear vs. Contract-Driven Parallelism. Unlike sequential workflows that suffer from context accumulation (Left), our Language Contract serves as a topological hypervisor, decoupling dependencies to enable parallel execution and global consistency via a hierarchical graph (Right).}
\label{fig:main}
\end{figure*}

Inspired by this separation of concerns, we propose Contract-Coding, a novel paradigm designed for Autonomous Symbolic Grounding. At its core is the \textit{Language Contract}, a formalized consensus that transforms vague "vibes" into a deterministic \textit{Single Source of Truth} (SSOT). Unlike transient prompt instructions, our Contract functions as a \textit{Symbolic Constraint Projection}~\cite{Neurosymbolic}: it autonomously compresses the high-dimensional intent space into orthogonal constraint subspaces. This decoupling allows us to model the repository as a Contract-Driven Hierarchical Graph (HEG). By using the Contract as a topological hypervisor, we reduce the informational dependency depth of tasks. This allows agents to execute in parallel, conditioned solely on the symbolic Contract rather than on the unstable implementation history, verified by a Contract-Guided Auditing mechanism.

Our contributions are threefold:
\begin{itemize} [itemsep=1pt, topsep=2pt, parsep=0pt]
\item We formalize the Language Contract mechanism, which autonomously operationalizes high-level user intents into structured symbolic paradigm, bridging the gap between ambiguous Vibe Coding and rigorous Spec-Coding.
\item We develop the Contract-Driven Hierarchical Graph paradigm. By utilizing the Contract as a semantic basis, we decrease informational dependency depth, unlocking massive \textit{Architectural Parallelism} as a natural consequence of symbolic decoupling.
\item We demonstrate that our approach has a great academic evolvement in greenfield repository generation. Crucially, we identify a Sub-linear Context Scaling effect: while repository scale expands significantly, the symbolic contract exhibits sub-linear growth, effectively compressing the reasoning context for complex systems.
\end{itemize}

\section{Related Work}

\paragraph{LLMs for Code and Autonomous Agents.}
Large Language Models (LLMs) have demonstrated profound capabilities in software engineering \citep{GPT-4, StarCoder, CodeLlama, DeepSeek-Coder, Lemur}. Beyond solitary code generation, the field has pivoted toward autonomous agents capable of planning and tool use \citep{xi2023risepotentiallargelanguage, Wang_2024, shen2023hugginggpt, patil2023gorilla}. Representative systems include general-purpose solvers like Auto-GPT \citep{Auto-GPT} and ToolFormer \citep{Toolformer}, as well as verticalized coding agents such as CodeAgent \citep{CodeAgent}, ToolCoder \citep{ToolCoder}, OpenHands \citep{OpenHands}, and SWE-agent \citep{SWE-agent}, which primarily focus on local bug-fixing or single-file synthesis.

\paragraph{Multi-Agent Collaboration Topologies.}
To mitigate the hallucination and context constraints of single agents, multi-agent systems (MAS) have been widely explored \citep{Multi-AgentColla, li2023camel}. Early frameworks such as MetaGPT \citep{MetaGPT} and ChatDev \citep{ChatDev} adopt linear ``waterfall'' workflows, effectively acting as a sequential Chain-of-Thought. More recent works emphasize dynamic or adaptive interaction structures \citep{AgentPrune, AutoFlow, ICLR_FLOW}, including SPP \citep{SPP}, AgentVerse \citep{AgentVerse}, DyLAN \citep{DyLAN}, and PriorDynaFlow \citep{PriorDynaFlow}. However, these systems still rely on conversational routing, where information propagates linearly, leading to the ``Context Exhaustion'' bottleneck identified in our theoretical analysis.

\paragraph{Spec-Driven Synthesis vs. Contract-Driven Orchestration.}
Unlike specification-driven methods~\citep{SpecAgent, ProjectCode} that treat constraints as static priors~\citep{ProtocolCoding}, Contract-Coding targets the spec-less ``Vibe Coding'' setting. We distinguish ourselves by (1) \textbf{Genesis}: autonomously synthesizing the contract from ambiguous intent rather than requiring formal inputs; and (2) \textbf{Topology}: elevating the contract to an active \textit{Hypervisor}. (3) \textbf{Information Hiding}: utilizing the Contract as a semantic barrier. This \textbf{compresses} the reasoning scope effectively preventing the context saturation typical of sequential baselines.

\section{Decoupling via Contract Constraints}
\label{sec:theory}

In this section, we formalize the repo-level generation task and analyze the probabilistic failure modes of existing paradigms. We demonstrate that the prevailing sequential approach suffers from inherent entropic divergence, which we resolve via a latent symbolic formulation.

\subsection{Task Formulation}
We define the repository generation task as maximizing the likelihood of synthesizing a valid code repository $\mathcal{R} = \{f_1, f_2, ..., f_N\}$ given a high-level, under-specified user intent $\mathcal{I}$:
\begin{equation}
\mathcal{R}^* = \operatorname*{arg\,max}_{\mathcal{R}} P(\mathcal{R} \mid \mathcal{I})
\end{equation}
A valid $\mathcal{R}$ requires not only local syntactical correctness but also global \textit{Symbolic Consistency}—meaning that any symbol (e.g., class `Player`) defined in file $f_i$ must be correctly referenced by $f_j$, regardless of their generation order.

\subsection{The Sequential Bottleneck}
\label{sec:sequential_bottleneck}

Standard multi-agent frameworks (e.g., MetaGPT, ChatDev) adopt a ``Chain-of-Thought'' topology. They approximate the joint probability using the probabilistic chain rule:
\begin{equation}
P(\mathcal{R} \mid \mathcal{I}) = \prod_{i=1}^{N} P(f_i \mid f_{<i}, \mathcal{I})
\end{equation}
where $f_{<i}$ represents the cumulative context history. While theoretically sound, this formulation introduces two systemic failures in the context of LLMs:

\paragraph{Sequential Error Propagation.} 
Since $f_i$ is conditioned on the raw implementation of $f_{<i}$, any latent defect $\epsilon$ in an early module $f_j (j \ll i)$ becomes part of the ground truth for all subsequent generation. The error term accumulates multiplicatively, causing the system to diverge from $\mathcal{I}$ toward a locally coherent but globally invalid state.

\paragraph{Context Exhaustion (The Signal-to-Noise Ratio).} 
As $N$ grows, the implementation context $f_{<i}$ expands linearly, inevitably exceeding the model's effective attention span. The informational density of the original intent $\mathcal{I}$ becomes diluted by verbose code details, leading to Symbolic Hallucination, where agents invent non-existent APIs to satisfy immediate local constraints.

\subsection{Conditional Independence via Symbolic Grounding}
\label{sec:resolution}

To break this curse of dimensionality, we reformulate the generation process by introducing the Language Contract $\mathcal{C}$ as a discrete \textit{Latent Symbolic Variable}. 

Instead of treating $\mathcal{C}$ as mere documentation, we posit it as the Information Barrier between architectural intent and implementation details. The joint probability is re-factorized as:
\begin{equation}
P(\mathcal{R} \mid \mathcal{I}) \approx P(\mathcal{C} \mid \mathcal{I}) \prod_{i=1}^{N} P(f_i \mid \mathcal{C})
\label{eq:parallel}
\end{equation}
This formulation introduces a fundamental topological shift:

\paragraph{Formalism 1 (Implementation Independence).}
Let $\mathcal{R}$ be the repository and $\mathcal{C}$ be the synthesized contract. Under the condition that $\mathcal{C}$ captures the sufficient interface semantics of $\mathcal{R}$(Guaranteed by the two-stage and audit mechanism), the implementation of any module $f_i$ is conditionally independent of the raw implementation of other modules $\mathcal{R}_{\setminus i}$:
\begin{equation}\label{para:form1}
    P(f_i \mid \mathcal{R}_{\setminus i}, \mathcal{C}) = P(f_i \mid \mathcal{C})
\end{equation}
This implies two critical properties:
\begin{enumerate}[itemsep=1pt, topsep=2pt, parsep=0pt]
    \item \textbf{Decoupled Execution:} The generation of $f_i$ depends solely on $\mathcal{C}$, not on the $O(N)$ volume of other source files.
    \item \textbf{Sub-linear Context Scaling:} While strict independence assumes an immutable $\mathcal{C}$, in practice, $\mathcal{C}$ serves as a compressed "Global View." Our empirical observations (Figure \ref{fig:token_entropy}) indicate that the symbolic tokens $|\mathcal{C}|$ grow sub-linearly relative to implementation tokens $|\mathcal{R}|$. Thus, agents maintain global awareness without suffering from context exhaustion.
\end{enumerate}

\section{The Contract-Coding Paradigm}
\label{sec:methodology}

Building upon the latent variable formulation, we implement \textsc{Contract-Coding}. As illustrated in Figure~\ref{fig:main}, the framework operationalizes the transition from vague intent to verified code through three distinct phases: (1) Symbolic Projection, (2) Topological Orchestration, and (3) Active Contract Auditing.

\subsection{Symbolic Projection}
\label{sec:phase1}

The objective is to synthesize a high-fidelity initial Contract $\mathcal{C}_0$ from intent $\mathcal{I}$. To bridge the gap between human-readable requirements and machine-executable constraints, we formalize this process as a Discrete Symbolic Evolution (DSE) over a hierarchical state space.

\paragraph{Hierarchical Contract Definition.} 
We define the Language Contract $\mathcal{C}$ as a dual-layer entity, acting as the interface between the neural agents and the symbolic system:
\begin{itemize}[itemsep=1pt, topsep=2pt, parsep=0pt]
    \item \textbf{The Constraint Projection (Structure):} Physically, $\mathcal{C}$ is structured as a collection of disjoint sections $\mathcal{C} = \{S_{\textsc{Req}}, S_{\textsc{Api}}, \dots\}$, organized into Product and Technical tiers. These sections provide the high-dimensional natural language context necessary for LLM reasoning.
    \item \textbf{The Executable Kernel (Symbolic):} For topological orchestration, we project the API Specification section ($S_{\textsc{Api}}$) into a rigorous logical kernel $\mathcal{K} = \langle \mathcal{N}, \Sigma, \Delta \rangle$. Here, $\mathcal{N}$ maps functional modules to file paths, $\Sigma$ defines strict type signatures, and $\Delta$ represents the state space.
\end{itemize}
This distinction is vital: while agents read the \textit{Constraint Projection} to understand "why", the execution graph (HEG) operates solely on the \textit{Executable Kernel} to determine "when" and "how".

\paragraph{Discrete Mutation Primitives.}
To prevent the semantic drift inherent in free-form text generation, we restrict the construction of $\mathcal{C}$ to a set of atomic mutation primitives $\mathbb{O} = \{\textsc{Add}, \textsc{Update}\}$. 
At any step $t$, an agent issues an action $a_t=\langle \text{op}, \text{sec}, \delta \rangle$, where $\text{sec} \in \mathcal{C}$ is the target section and $\delta$ is the structured content. The transition function applies this mutation: $\mathcal{C}_{t+1} = T(\mathcal{C}_t, a_t)$. Crucially, any change to the text triggers an immediate re-projection of the kernel $\mathcal{K}$. If an action results in a kernel that violates graph acyclicity or type safety, the system intercepts and rejects it before it pollutes the global state.

\paragraph{Two-Stage Initialization Protocol.}
Leveraging the DSE mechanism, we mitigate "one-shot" hallucinations via a streamlined pipeline:
\begin{enumerate}[itemsep=1pt, topsep=2pt, parsep=0pt]
    \item \textbf{Proposal (Generator):} The Generator Agent analyzes $\mathcal{I}$ and emits a sequence of mutation actions to construct the preliminary contract $\mathcal{C}_{draft}$. This effectively discretizes the continuous Vibe manifold into the structured state space.
    \item \textbf{Rectification (Discriminator):} The Discriminator Agent performs a one-pass audit on $\mathcal{C}_{draft}$. It verifies architectural soundness (e.g., ensuring $\Delta$ is connected and acyclic) and enforces a \textit{Completeness Constraint}—rejecting any module description in $\mathcal{K}$ that lacks sufficient detail. 
\end{enumerate}
This phase may be imperfect. Residual ambiguities are propagated to the \textit{Topological Orchestration} phase, where they are resolved dynamically by downstream execution agents (e.g., Code Reviewer) via runtime `Update' actions to the Contract.

\begin{figure}[t]
  \centering
  \includegraphics[width=\linewidth]{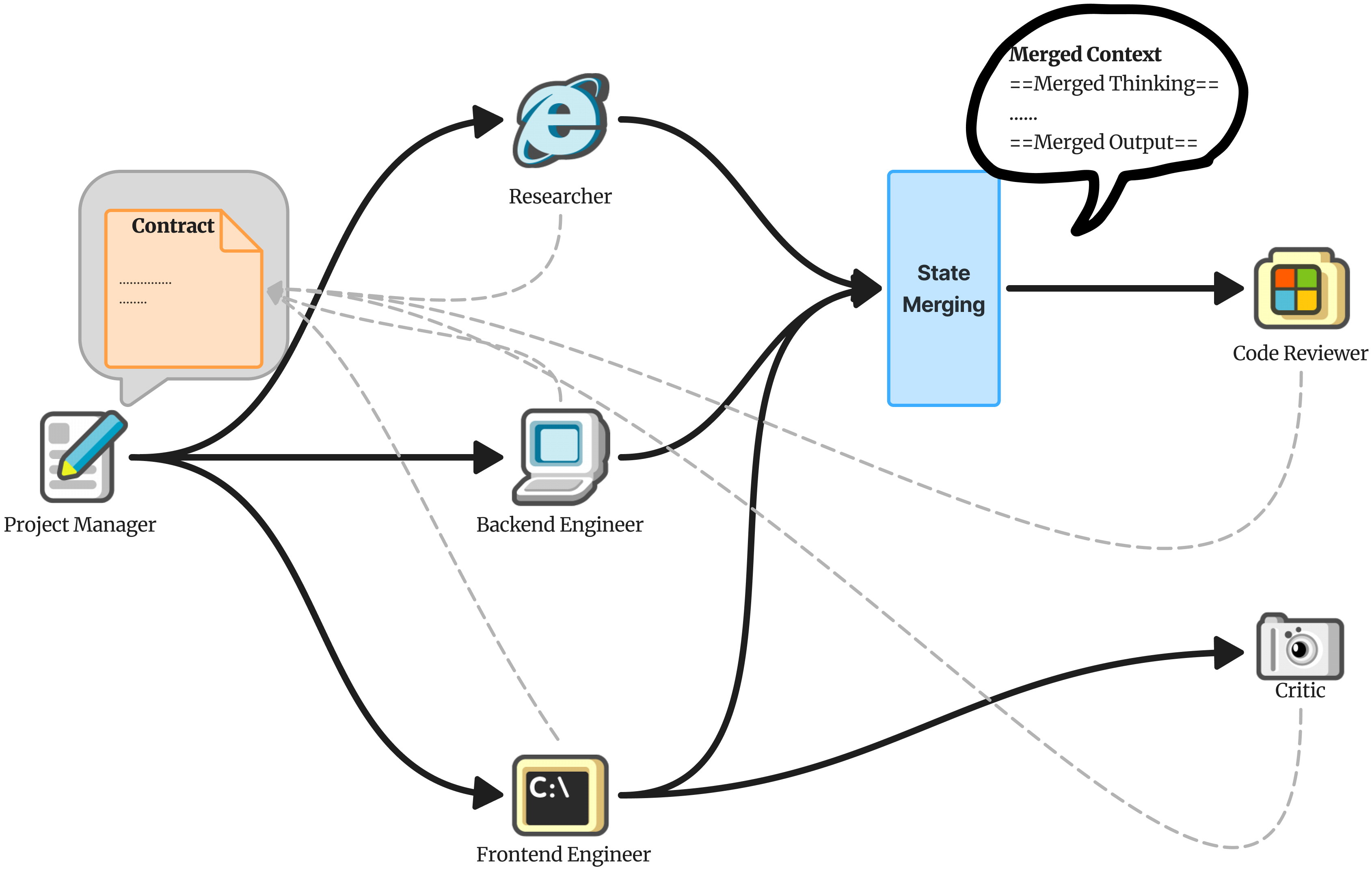}
  \caption{Contract-driven execution. The Contract defines the global state matrix, which drives the Hierarchical Execution Graph (HEG) to schedule parallel agents.}
  \label{fig:layer}
\end{figure}

\subsection{Topological Orchestration (The HEG)}
\label{sec:heg}

Traditional multi-agent systems often rely on hard-coded chains. In contrast, we propose a Dynamic Hierarchical Execution Graph (HEG, Figure~\ref{fig:layer}), where the execution topology $G^{(t)}$ is strictly derived from the Contract.

Let $\mathcal{T} = \{\tau_1, \dots, \tau_K\}$ be the set of atomic tasks defined in $\mathcal{C}$. Each task $\tau_i$ possesses a lifecycle status $\sigma_i \in \{\textsc{Todo(T)}, \textsc{Done(D)}, \textsc{Error(E)}, \textsc{Verified(V)}\}$. The HEG functions as a state-conditional scheduler $\Phi$:
\begin{equation}
    \Phi(\tau_i) = 
    \begin{cases} 
      \text{Worker}(\tau_i, \mathcal{C}) & \text{if } \sigma_i \in \{\textsc{T}, \textsc{E}\} \\
      \text{Verifier}(\tau_i, \mathcal{C}, \mathcal{I}_{impl}) & \text{if } \sigma_i = \textsc{D} \\
      \emptyset & \text{if } \sigma_i = \textsc{V}
    \end{cases}
\end{equation}
This mapping induces three distinct operational modes:

\paragraph{Generative Expansion (Parallel Implementation).} 
For tasks marked as \textsc{Todo} or \textsc{ERROR}, the system instantiates a worker node. Crucially, the input context provided to the worker is $\langle \tau_i, \mathcal{C} \rangle$. By providing the Full Contract $\mathcal{C}$ rather than raw code, we enable the agent to understand global dependencies while avoiding the noise of implementation details. This validation Formalism~\ref{para:form1} allows multiple workers to execute in parallel without lock-step synchronization.

\paragraph{Critical Evaluation (Normative Verification).} 
When a task transitions to \textsc{Done}, the graph spawns a \textit{Critic} node. This agent receives the implementation $\mathcal{I}_{impl}$ and validates it against the Contract $\mathcal{C}$. The objective is strictly normative: determining whether the code fulfills the contract signature, not whether it "looks good."

\paragraph{Convergence.} 
The process iterates until $\forall \tau_i, \sigma_i = \textsc{Verified}$. This design eliminates explicit "dispatch" instructions; the workflow is an emergent property of the repository's state resolution. Our empirical observation is that the \textit{Critic Evaluation} efficiently resolve most of the issues. As a final safety mechanism to ensure decidability, we can impose a Maximum Topological Depth $T_{max}$. If the system fails to converge within $T_{max}$ layers, it terminates gracefully, yielding the current best-effort repository. This ensures that the system is bounded and avoids infinite resource consumption.

This validation of Formalism~\ref{para:form1} allows multiple workers to execute in parallel without lock-step synchronization. The complete orchestration procedure, including the state transitions and parallel dispatching, is formalized in Algorithm~\ref{alg:auditing}.

\subsection{Active Contract Auditing}
\label{sec:auditing}

To close the loop between execution and state, we introduce a deterministic Contract Auditor. Unlike passive loggers, the Auditor functions as the system's Homeostatic Controller. It continuously monitors the graph state and triggers \textit{Topological Interventions} when deviations are detected.

\paragraph{Structural Alignment ($E(\mathcal{C})$).} 
This metric enforces that the implementation physically aligns with the Contract. Let $U$ be the set of existing file units.
\begin{equation}
    E(\mathcal{C}) = \frac{1}{|\mathcal{T}|} \sum_{\tau \in \mathcal{T}} \mathbb{I}(\exists u \in U : \text{match}(\tau, u))
\end{equation}
\textbf{Intervention:} If $E(\mathcal{C})$ fails (i.e., a required symbol is missing), the Auditor performs Task Injection. It dynamically inserts a new implementation node into the HEG for the next time step, forcing the system to fill the "Hollow Skeleton."

\paragraph{State Synchronization ($S(\mathcal{C})$).} 
The Auditor synchronizes high-level planning with low-level execution outcomes by parsing agent logs.
\begin{equation}
   S(C)=\{ \sigma_i^{(t+1)} | \sigma_i^{(t+1)} \leftarrow \text{Audit}(\sigma_i^{(t)}, \text{Output}_{\mathcal{A}}) \}
\end{equation}
\textbf{Intervention:} A failure here triggers Status Regression. For example, if a \textit{Critic} agent flags a logic bug, the Auditor regresses the task status from \textsc{Done} to \textsc{Error}. This automatically triggers the HEG to reschedule the owner agent for repair.

\paragraph{Consistency Control ($V(\mathcal{C})$).} 
Consistency is defined as the alignment between the Contract's recorded signatures and the actual workspace code:
\begin{equation}
    V(\mathcal{C}) = \bigwedge_{\tau \in \mathcal{T}} (\text{state}_{\mathcal{C}}(\tau) \equiv \text{state}_{W}(\tau))
\end{equation}
\textbf{Intervention:} If a mismatch is detected (e.g., an agent silently modifies a function signature), the Auditor enforces Normative Alignment. It rejects the invalid state transition and injects a \textit{Synchronization Task}, compelling the agent to either revert the code or formally propose a Contract Amendment (Update Action) to legitimize the change.

\begin{figure*}[!htbp]  
  \centering  
  \includegraphics[width=\linewidth]{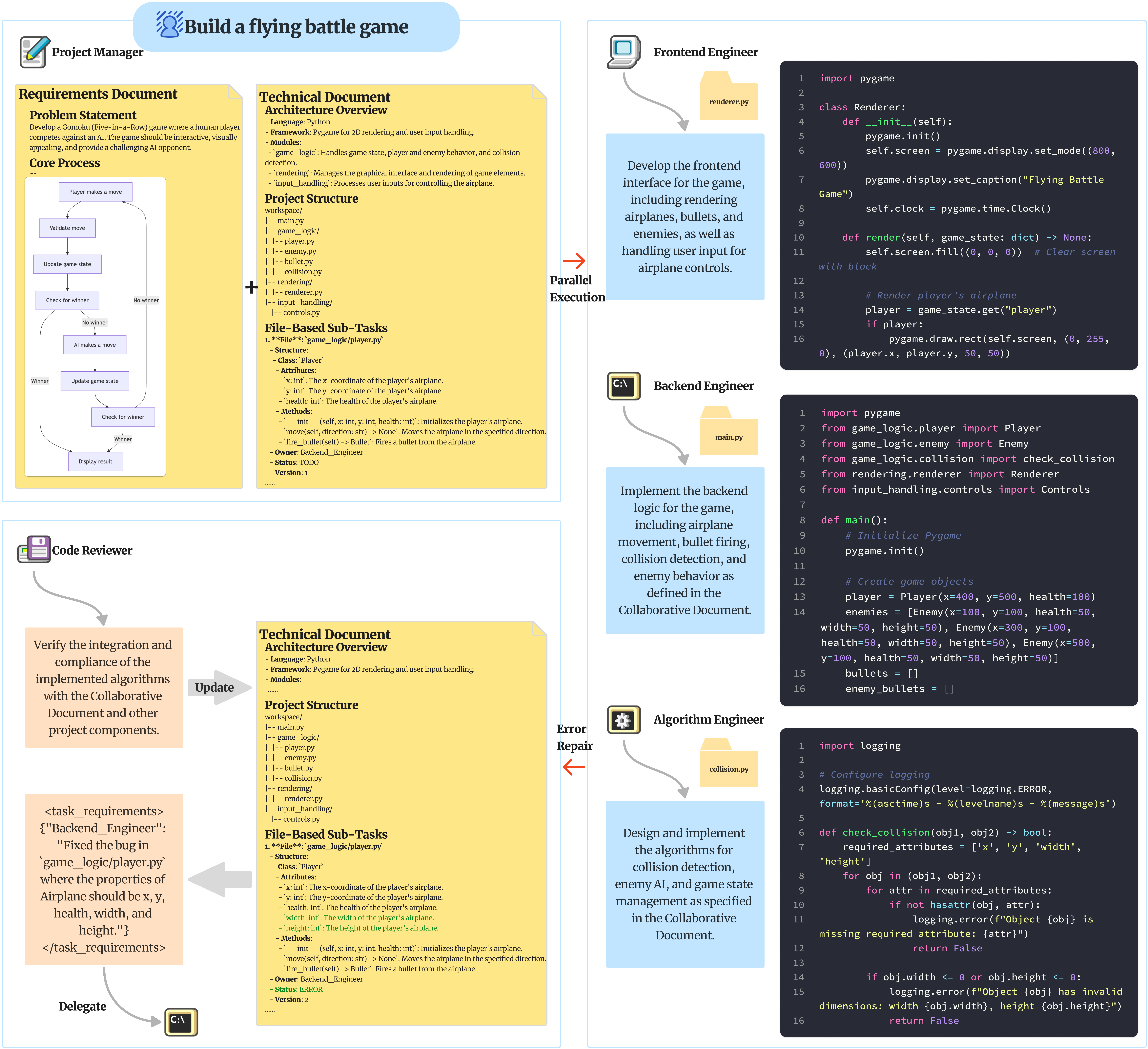}  
  \caption{Contract-Guided Self-Healing Case Study. When divergent assumptions (e.g., missing attributes) occur between parallel Backend/Algorithm agents, Contract Auditing detects schema violations during merging and triggers Normative Alignment to auto-correct semantic mismatches.}  
  \label{fig:details}  
\end{figure*}

\section{Experiments}
We evaluate \textsc{Contract-Coding} on a \textit{Complexity Spectrum} of five greenfield repositories, ranging from simple algorithmic scripts to complex, event-driven systems. Our primary goal is to verify if the proposed \textit{Symbolic Decoupling} can break the ``Inverse Scalability Law'' typically observed in autonomous agents.

\subsection{Experimental Setup}

\paragraph{Benchmarks: The Greenfield-5 Suite.}
We introduce Greenfield-5 to bridge the gap between algorithmic scripts and system-level architectures. The suite spans a complexity spectrum: (1) \textbf{Gomoku} (Logic); (2) \textbf{Plane Battle} \& \textbf{Snake++} (Event-Driven); (3) \textbf{City Sim} (Resource Mgmt, $\sim$11 files); and (4) \textbf{Roguelike} (Scalability, 15--25 files). All tasks mandate a strict multi-file structure to verify architectural integrity. Full specifications are in Appendix~\ref{app:benchmark}.

\paragraph{Baselines.}
We compare against: (1) \textbf{Commercial AI IDEs} (Lingma, Trae, Gemini Studio) serving as upper-bound references; and (2) \textbf{Academic Frameworks} (MetaGPT, ChatDev, FLOW, OpenHands) representing sequential multi-agent paradigms.

\paragraph{Metrics.}
We report "Success Rate" (a composite of Executability, Interactivity, and Rule Adherence) and "Efficiency" (Wall-clock Time). Additionally, "File Count" tracks code bloat. Results represent the mean of 10 independent runs.

\paragraph{Ablations.}
To validate our topological contributions, we compare the full model against: (1) \textbf{w/o HEG} (Sequential execution); (2) \textbf{w/o Contract} (Removing the symbolic layer); and (3) \textbf{Model Agnosticism} (Replacing the backbone with Qwen-Plus). Detailed configurations are in Appendix~\ref{app:ablation}.

\subsection{Overall Success \& Scalability(RQ1)}

Table~\ref{tab:main_results} reports performance across the Greenfield-5 suite. We observe three key trends:

\paragraph{The "Complexity Wall" for Legacy Agents.}
While traditional multi-agent frameworks (MetaGPT, ChatDev, FLOW) perform adequately on logic-centric tasks (Gomoku), they suffer from \textit{Architectural Collapse} on multi-module challenges. Even the current SOTA OpenHands, despite achieving 100\% success on Gomoku, experiences a significant performance drop-off to 50\% on Snake++ and 30\% on Roguelike. This highlights their struggle with "Context Saturation" and managing deep inter-file dependencies inherent in repo-level synthesis.

\paragraph{Efficiency and Structural Rigor vs. Commercial SOTA.}
Commercial AI IDEs establish a strong upper bound, achieving high Overall Success rates through proprietary optimizations. Notably, Gemini Studio demonstrates competitive speed (e.g., 49.3s on City Sim) and success (63\% on Roguelike). We hypothesize this leverages its "ultra-long context window", enabling a single powerful model to process extensive cross-file information, thus effectively circumventing traditional linear reasoning bottlenecks. However, even this brute-force context scaling shows diminishing returns, with Gemini's performance declining on more complex tasks. Other commercial tools, like Trae, often incur extreme latency ($>350$s) and significant code bloat (Files=19.0).

\paragraph{Limits of Parallelism.}
On the scalability stress-test (Roguelike), our fully parallel approach sees a performance drop (47\% success rate) compared to the SOTA (CodeBuddy), reflecting the challenge of integrating 25+ coupled files without intermediate feedback. 
However, distinct from the "Silent Failures" in legacy baselines, our failures are mostly localized logic errors. A detailed forensic analysis of failure modes across all methods (e.g., MetaGPT's "Hollow Skeleton" vs. Trae's "Logical Detachment") is provided in Appendix~\ref{sec:app_failure_analysis}.

\begin{table*}[t]
\centering
\setlength{\tabcolsep}{3.5pt} 
\renewcommand{\arraystretch}{0.9}
\resizebox{\textwidth}{!}{%
\begin{tabular}{l|ccc|ccc|ccc|ccc|ccc}
\toprule
\multirow{2}{*}{\textbf{Method}} & \multicolumn{3}{c|}{\textbf{Gomoku}} & \multicolumn{3}{c|}{\textbf{Plane Battle}} & \multicolumn{3}{c|}{\textbf{City Sim}} & \multicolumn{3}{c|}{\textbf{Snake++}} & \multicolumn{3}{c}{\textbf{Roguelike}} \\
 & \small{Overall} & \small{Files} & \small{Time} & \small{Overall} & \small{Files} & \small{Time} & \small{Overall} & \small{Files} & \small{Time} & \small{Overall} & \small{Files} & \small{Time} & \small{Overall} & \small{Files} & \small{Time} \\ 
\midrule
\multicolumn{16}{c}{\textit{Legacy Multi-Agent Frameworks}} \\
\midrule
MetaGPT & 70 & 5.2 & 126s & 50 & 4.0 & 129s & 30 & 3.0 & 185s & 10 & 4.0 & 210s & 10 & 10.9 & 261s \\
ChatDev & 93 & 4.4 & 88s & 90 & 5.1 & 79s & 45 & 6.0 & 140s & 25 & 7.2 & 165s & 10 & 13.9 & 144s \\
FLOW & 96 & 1.0 & 155s & 76 & 1.0 & 76s & 60 & 1.0 & 110s & 30 & 1.0 & 130s & 0 & 1.0 & 90s \\ 
OpenHands & 100 & 3.3 & 173s & 90 & 4.9 & 182s & 63 & 7.3 & 296s & 50 & 13.7 & 267s & 30 & 14.1 & 311s \\
\midrule
\multicolumn{16}{c}{\textit{Commercial AI IDEs (SOTA)}} \\
\midrule
Lingma & \textbf{100} & 1.0 & 127s & 86 & 1.4 & 125s & 75 & 8.0 & 220s & 60 & 12.0 & 310s & 33 & 24.7 & 620s \\
Trae & 70 & 13.7 & 393s & 80 & 14.5 & 411s & 85 & 19.0 & 350s & 65 & 22.0 & 412s & 47 & 23.9 & 218s \\
Gemini Studio & 100 & 10.3 & 103s & 100 & 9.6 & 165s & 87 & 13.0 & 49s & 83 & 15.5 & 108s & 63 & 16.6 & 72.1s \\
CodeBuddy & 96 & 2.8 & 91s & \textbf{100} & 3.0 & 92s & 60 & 14.0 & 196s & 60 & 11.7 & 414s & 40 & 19.6 & 422s \\ 
\midrule
\multicolumn{16}{c}{\textit{Ours (Autonomous)}} \\
\midrule
Ours w/o HEG & \textbf{100} & 4.0 & 205s & \textbf{100} & 6.1 & 201s & 85 & 11.2 & 480s & 78 & 16.0 & 465s & 47 & 25.0 & 510s \\ 
\textbf{Ours (Full)} & \textbf{100} & 4.0 & 136s & \textbf{100} & 6.2 & 117s & 87 & 11.2 & 257s & \textbf{80} & 16.2 & 198s & 47 & 14.2 & 232s \\ 
\bottomrule
\end{tabular}%
}
\caption{\textbf{Main Results on Repository-Level Generation.} Comparison across five tasks of decreasing complexity. \textbf{Overall}: Average of Executability, Interactivity, and Rule Adherence. \textbf{Files}: Generated file count (closer to ground truth is better). \textbf{Time}: End-to-end latency.}
\label{tab:main_results}
\end{table*}

\subsection{The Context Scaling Effect(RQ2)}

\begin{figure}[t]
\centering
\includegraphics[width=\linewidth]{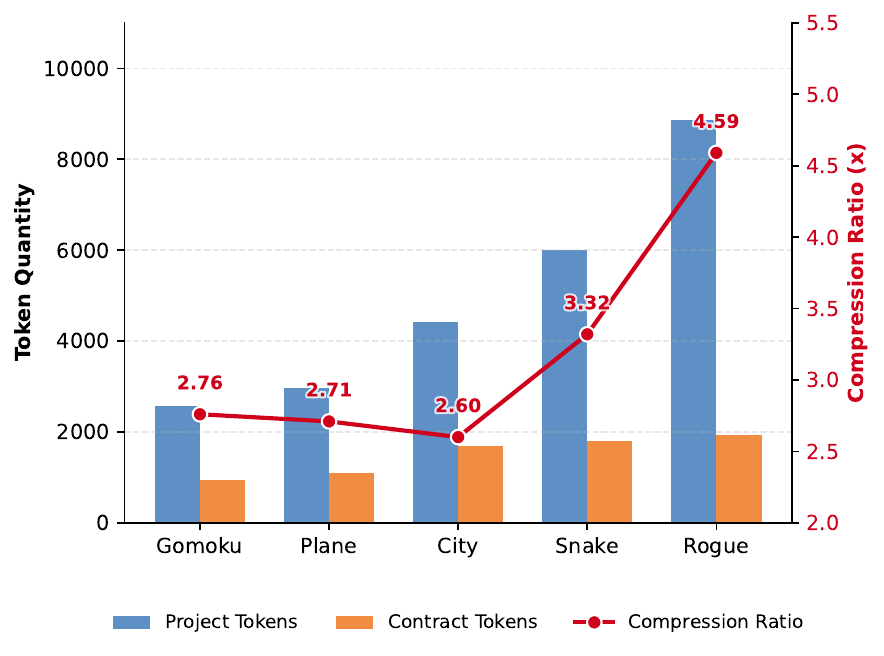}
\caption{\textbf{Token Decoupling Analysis.} As project complexity grows, the Language Contract size remains stable. Red markers show the compression efficiency.}
\label{fig:token_entropy}
\end{figure}

Is the efficiency gain purely coincidental? Figure~\ref{fig:token_entropy} illustrates the token dynamics across the five tasks. 

As the project complexity grows ($T_{proj}$), the size of the Language Contract ($T_{cont}$) grows at a much slower rate. This confirms that our \textit{Constraint Projection} mechanism effectively compresses the high-dimensional intent space. For the \textit{Roguelike} challenge (8,857 tokens), our agents only need to attend to a $\sim$1,900-token contract, avoiding the ``Lost-in-the-Middle'' phenomenon plaguing standard RAG or long-context approaches. This 4.6x compression effectively downgrades a repository-level task to a series of manageable single-file tasks.

\subsection{Case Study: Conflict Repair(RQ3)}
To demonstrate the system's robustness against initial contract ambiguities and logical errors, we analyze a real-world Conflict-Repair cycle from the Plane Battle task (visualized in Figure \ref{fig:details}). During the parallel execution phase, a divergence occurred due to under-specified requirements: the Algorithm Engineer, implementing collision logic, implicitly assumed the existence of \texttt{width} and \texttt{height} attributes in the \texttt{Player} class. Conversely, the Backend Engineer, adhering strictly to the initial Contract, implemented only basic attributes (\texttt{x}, \texttt{y}, \texttt{health}).

This divergence was autonomously resolved via the system mechanism. The Code Reviewer detected a semantic mismatch between the Algorithm's invocation (demand) and the Backend's definition (supply). Crucially, recognizing this as an incorrectness in the Contract itself rather than a mere coding error, the Code Reviewer retroactively patched the Contract definition, formally adding the spatial dimensions to the schema as the new Single Source of Truth. The task was flagged as \texttt{ERROR}, triggering a targeted re-dispatch to the Backend Agent with the directive: ``\textit{Fix the schema definition; Player requires spatial dimensions...}'' This highlights the systemic robustness of our paradigm: by synergizing the Language Contract with reflective auditing, the framework effectively prevents error propagation of early logical flaws. This dynamic self-healing capability serves as the decisive factor enabling our high task overall success rate, ensuring that local schema inconsistencies are resolved internally rather than propagating into global system failures.

\subsection{Scalability Analysis: The Roguelike Challenge (RQ4)}
\label{sec:rq4}
To probe the upper limits of autonomous architecture, we introduce a complex Roguelike task requiring 15-25 interdependent files. Table~\ref{tab:basic_tasks_detail} highlights the divergence between paradigms.

\paragraph{Systemic Collapse in Academic Baselines.} 
Current academic models exhibit severe architectural decoupling at scale. As detailed in Appendix~\ref{sec:app_failure_analysis}, MetaGPT and ChatDev suffer from \textit{Implementation Sparsity} and \textit{Grounding Failures}, respectively, while FLOW faces \textit{Context Collapse}, rendering them unable to synthesize valid multi-file projects. OpenHands may encounter \textit{Lost-in-the-Middle}.

\paragraph{Commercial SOTA: The Resource-Efficiency Trade-off.} 
Commercial IDEs (e.g., Gemini Studio) establish a strong upper bound (63\% success). However, this relies on massive context windows ($>$200k tokens) and proprietary infrastructure. In stark contrast, our experiments were strictly confined to a 16k token limit. That \textsc{Contract-Coding} achieves 100\% Structural Integrity under this constraint validates our core hypothesis: \textit{Architectural Decoupling is a more scalable path than Brute-Force Context Scaling.} Our method effectively "simulates" infinite context via symbolic compression, whereas other models (e.g., Gemini, Lingma) suffer from \textit{Probabilistic Drift} or infinite recursive loops when context overflows.

\paragraph{Architectural Integrity over Algorithmic Fidelity.} 
Although our method lags in absolute overall success (47\%), the nature of failure is distinct. Our failures are strictly confined to "Local Logic Errors" (e.g., suboptimal heuristics) driven by base model limitations, not architectural collapse. This distinction is critical for ASE: we prioritize Structure over Detail. In real-world workflows, fixing local bugs within a perfect architecture is trivial; refactoring a "working" but structurally chaotic codebase is prohibitively expensive.

\section{Conclusion}

Contract-Coding, with Language Contract as SSOT, decouples architectural complexity from implementation details and enable efficient parallel execution. Validation on the Greenfield-5, compared with SOTA baselines, this method achieves an up to 4.6× token compression ratio as well as an overall improvement in success rate. Our work concludes that more complex vibe coding can be realized through auditable structural-symbolic contracts, instead of relying on larger context windows.


\section{Limitations}
Despite the performance gains, our framework has several limitations that provide avenues for future research.First, our claim of $O(1)$ dependency depth pertains to the \textit{informational topology} during parallel implementation. However, the total system latency is subject to the synchronization-repair rounds ($R$). In scenarios with extreme inter-module semantic coupling, $R$ may scale with repository complexity, potentially converging toward a serialized bottleneck.

Second, regarding Benchmark Alignment, we observe a significant gap between existing evaluation suites and the requirements of autonomous repository synthesis. Current benchmarks like SWE-bench \cite{SWE-agent} primarily focus on \textit{Brownfield Repair} (i.e., surgical bug-fixing in mature codebases). In contrast, our work targets Greenfield Synthesis from high-level intents (often referred to as "vibe coding" in developer communities), which emphasizes architectural orchestration and cross-module consistency. Due to the lack of standardized benchmarks for multi-file generative synthesis, we utilized curated game-engine tasks. We are committed to developing a specialized, automated benchmark in future work to formally evaluate the Scaling-up laws of contract-driven paradigms across larger, heterogeneous software systems ($>100$ files).

Finally, our baseline comparisons involve commercial "black-box" IDEs; although we standardized the autonomous environment, their evolving internal models may introduce temporal variance in reproducibility.

\section{Ethical Considerations}
\label{sec:ethics}

\subsubsection{Dual-Use Risks and Safety.} The advancement of autonomous software engineering lowers the technical barrier for software creation, which inevitably raises concerns regarding the rapid generation of malicious code (e.g., malware or automated hacking scripts). However, unlike end-to-end ``black box'' approaches that generate opaque deliverables, our \textit{Contract-Driven Paradigm} inherently promotes Accountability. By enforcing a human-readable \textit{Language Contract} as the intermediate control layer, our framework provides a mandatory auditing checkpoint, making it significantly easier to detect and intercept malicious intent before execution.

\subsubsection{Carbon Footprint.} We acknowledge that multi-agent frameworks involve redundant communication, leading to higher token consumption compared to single-pass generation. We advocate for the responsible use of such systems, recommending their deployment primarily for complex, architectural-level tasks where the high reliability justifies the computational cost.

\bibliography{custom}

\appendix

\section{Appendix: Implementation Details of Conflict Control}
\label{app:conflict}

While the Hierarchical Execution Graph enables parallel execution, the inherent asynchrony introduces the risk of ``lost updates.'' We mitigate this via a \textbf{Differential Interval Analysis} mechanism.

The system enforces a Synchronized Commit Protocol where updates from the current layer are aggregated only after all agents have completed tasks. We utilize the initial Contract state $C_{base}$ as the \textbf{Unique Coordinate System}. By calculating line offsets relative solely to this immutable baseline rather than dynamic intermediate states, we eliminate Positional Coupling (errors caused by index shifting).

\begin{table*}[t]
    \centering
    \small 
    \label{tab:ide_config}
    \begin{tabular}{lcccc}
        \toprule
        \textbf{AI IDE} & \textbf{Version} & \textbf{VSCode Ver.} & \textbf{Backbone Model} & \textbf{Mode} \\
        \midrule
        Lingma & 0.2.3 & 1.100.0 & Default & Agent \\
        CodeBuddy & 4.1.1 & 1.100.0 & GPT-5.1-Codex-Max & Craft \\
        Trae & 3.2.1 & 1.104.3 & GPT-5 Medium & Solo Build \\
        Gemini Studio & -- & -- & Gemini 3 Flash Preview & Build \\
        \bottomrule
    \end{tabular}
    \caption{\textbf{Configuration of Commercial AI IDE Baselines.} Detailed specifications of the IDE versions, underlying VSCode engines, and backbone models used in our comparative evaluation. All AI IDEs choose their corresponding Vibe coding mode.}
\end{table*}

\paragraph{Qualitative Analysis: Why Union-First?}
We deliberately employ a \textbf{Union-First Strategy} for resolving semantic conflicts in the Contract, as opposed to a naive ``Last-Write-Wins'' policy. 
\begin{itemize}
    \item \textbf{Information Preservation:} In LLM-generated content, different agents may discover orthogonal requirements for the same interface (e.g., Agent A adds `width`, Agent B adds `color`). A Last-Write-Wins policy would silently discard one agent's contribution, leading to downstream ``Variable Not Found'' errors.
    \item \textbf{Auditing Efficiency:} The Union-First approach preserves the superset of all proposed constraints. While this may introduce redundancy, removing redundant lines during the Audit phase (Reduction) is computationally easier and less prone to hallucination than re-generating missing logic from scratch (Synthesis).
\end{itemize}
We acknowledge that for extremely large teams, this strategy could lead to contract bloating. Future work will explore \textit{Conflict-Aware Merging} using a dedicated arbiter LLM to semantically deduplicate the union set.

\begin{figure}[h] 
  \centering 
  \includegraphics[width=\linewidth]{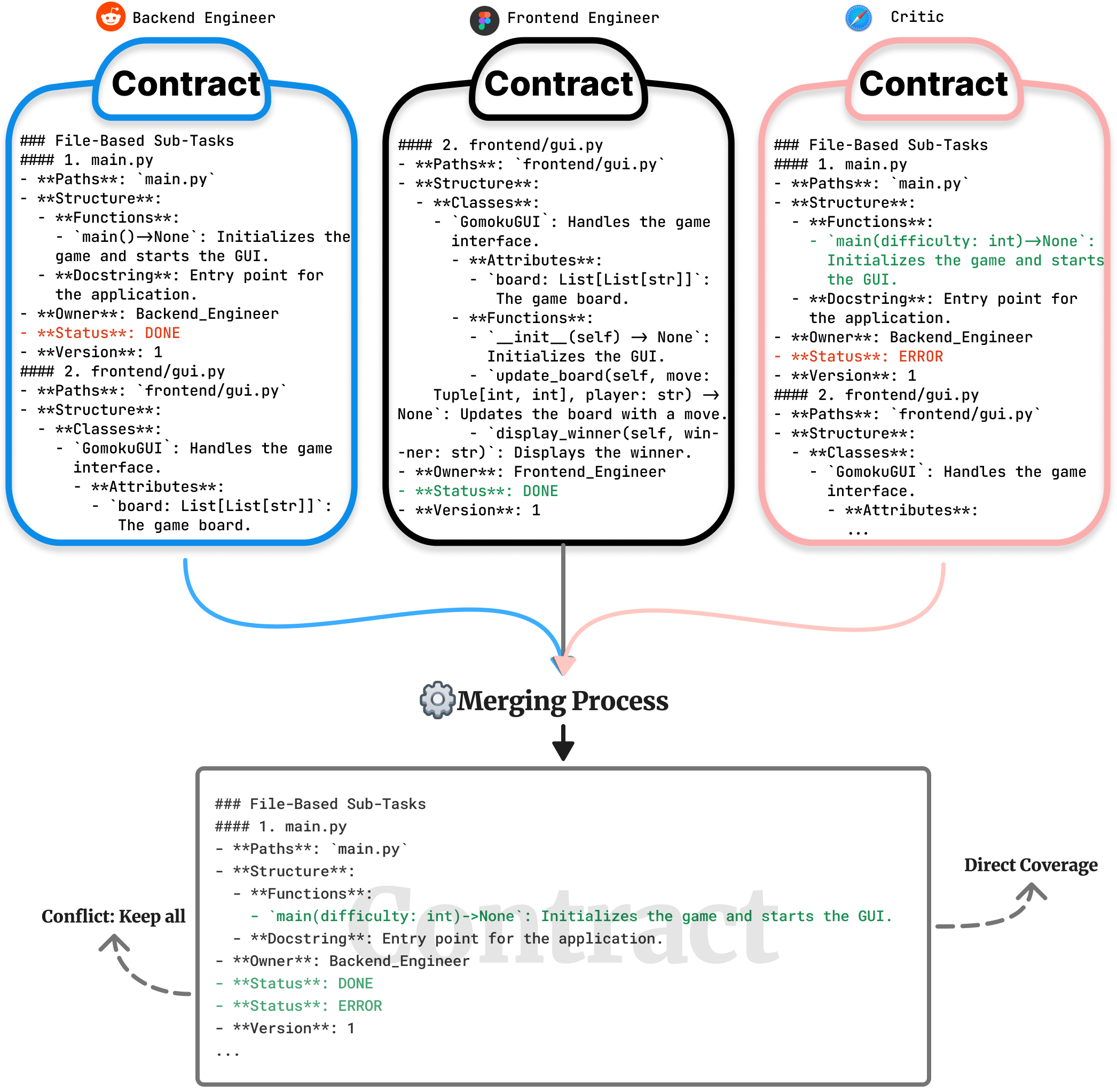}
  \caption{Conflict Control via Differential Interval Analysis. Using $C_{base}$ as an immutable baseline, we decompose updates into Atomic Patches. Semantic conflicts are resolved via a Union-First Strategy, prioritizing information preservation to prevent silent data loss.} 
  \label{fig:conflict} 
\end{figure}

\section{Appendix: Experimental Environment}
\label{app:setup}

\paragraph{Hardware Specifications.} 
Experiments were conducted on a MacBook Pro (M2 Pro) equipped with 32GB of unified memory and a 512GB SSD, running macOS 15. The explicit hardware constraint ensures that our efficiency metrics reflect performance on consumer-grade devices rather than high-end server clusters.

\paragraph{Hyperparameters.} 
Except for the model ablation experiment, all academic method experiments utilized \texttt{gpt-4o-2024-11-20} with a temperature of 0.0 to ensure determinism and a context window limit of 16k tokens. The specific versions and configurations of the commercial AI IDE baselines are detailed in Table~\ref{tab:ide_config}.

\begin{table*}[t]
\centering
\setlength{\tabcolsep}{4pt} 
\small 
\begin{tabular}{l|ccccc c|ccccc c|c}
\toprule
 & \multicolumn{6}{c|}{\textbf{Task 1: Gomoku (Logic)}} & \multicolumn{6}{c|}{\textbf{Task 2: Plane Battle (Action)}} & \\
\cmidrule(lr){2-7} \cmidrule(lr){8-13}
\textbf{Method} & \textbf{Exec} & \textbf{Inter} & \textbf{Rule} & \textbf{Succ.} & \textbf{Time(s)} & \textbf{Files} & \textbf{Exec} & \textbf{Inter} & \textbf{Rule} & \textbf{Succ.} & \textbf{Time(s)} & \textbf{Files} & \textbf{Parallel} \\ 
\midrule
\multicolumn{14}{c}{\textit{Legacy Multi-Agent Frameworks}} \\
\midrule
MetaGPT       & 70  & 70  & 70  & 70\%  & 126.0 & 5.2 & 50  & 50  & 50  & 50\%  & 128.5 & 4.0 & $\times$ \\
ChatDev       & 100 & 90  & 90  & 93\%  & \textbf{87.9} & 4.4 & 90  & 90  & 90  & 90\%  & 79.4  & 5.1 & $\times$ \\
FLOW          & 100 & 90  & 100 & 97\%  & 155.0 & 1.0 & 100 & 80  & 50  & 77\%  & \textbf{75.7}  & 1.0 & \checkmark \\
OpenHands     & 100 & 100 & 100 & 100\% & 173.2 & 3.3 & 100 & 100 & 100 & 100\% & 182.0 & 4.9 & $\times$ \\
\midrule
\multicolumn{14}{c}{\textit{Commercial AI IDEs}} \\
\midrule
Lingma        & 100 & 100 & 100 & 100\% & 127.0 & 1.0 & 90  & 90  & 80  & 83\%  & 125.0 & 1.4 & $\times$ \\
Trae          & 70  & 70  & 70  & 70\%  & 393.0 & 13.7& 80  & 80  & 80  & 80\%  & 411.0 & 14.5& $\times$ \\
Gemini Studio & 100 & 100 & 100 & 100\%  & 103.0 & 10.3& 100 & 100 & 100 & 100\% & 165.0 & 9.6 & $\times$ \\
CodeBuddy     & 100 & 90  & 100 & 93\%  & 91.0  & 2.8 & 100 & 100 & 100 & 100\% & 92.0  & 3.0 & $\times$ \\
\midrule
\multicolumn{14}{c}{\textit{Ours (Autonomous)}} \\
\midrule
Ours w/o HEG  & 100 & 100 & 100 & 100\% & 205.0 & 4.0 & 100 & 100 & 100 & 100\%   & 201.0 & 6.1 & $\times$ \\
\textbf{Ours (Full)} & \textbf{100} & \textbf{100} & \textbf{100} & \textbf{100\%} & 136.0 & 4.0 & \textbf{100} & \textbf{100} & \textbf{100} & \textbf{100\%} & 117.0 & 6.2 & \checkmark \\ 
\bottomrule
\end{tabular}
\caption{\textbf{Detailed Metrics: Basic Tasks.} Comparison on Logic-Oriented (Gomoku) and Action-Oriented (Plane Battle) tasks. Gemini Studio demonstrates high speed but produces bloated file structures compared to our optimized output.}
\label{tab:basic_tasks_detail}
\end{table*}

\begin{table*}[t]
\centering
\label{tab:inter_tasks_detail}
\setlength{\tabcolsep}{4pt} 
\small
\begin{tabular}{l|cccccc|cccccc|c}
\toprule
 & \multicolumn{6}{c|}{\textbf{Task 3: City Sim (11 Files)}} & \multicolumn{6}{c|}{\textbf{Task 4: Snake++ (16 Files)}} & \\
\cmidrule(lr){2-7} \cmidrule(lr){8-13}
\textbf{Method} & \textbf{Exec} & \textbf{Inter} & \textbf{Rule} & \textbf{Succ.} & \textbf{Time(s)} & \textbf{Files} & \textbf{Exec} & \textbf{Inter} & \textbf{Rule} & \textbf{Succ.} & \textbf{Time(s)} & \textbf{Files} & \textbf{Parallel} \\ 
\midrule
\multicolumn{14}{c}{\textit{Legacy Multi-Agent Frameworks}} \\
\midrule
MetaGPT       & 50  & 20  & 20  & 20\%  & 185 & 3.0 & 20  & 10  & 0   & 0\%   & 210 & 4.0  & $\times$ \\
ChatDev       & 60  & 40  & 35  & 30\%  & 140 & 6.0 & 30  & 30  & 15  & 10\%  & 165 & 7.2  & $\times$ \\
FLOW          & 60  & 60  & 60  & 40\%  & 110 & 1.0 & 30  & 30  & 30  & 0\%   & 130 & 1.0  & \checkmark \\
OpenHands     & 100 & 40 & 50 & 63\% & 296.4 & 7.3 & 60 & 50 & 40 & 50\% & 266.8 & 13.7 & $\times$ \\
\midrule
\multicolumn{14}{c}{\textit{Commercial AI IDEs}} \\
\midrule
Lingma        & 90  & 70  & 70  & 77\%  & 446.8 & 10.6 & 80 & 50  & 50  & 60\% & 453.6 & 12.9 & $\times$ \\
Trae          & 100  & 90  & 90 & 93\% & 231.2 & 15.3 & 90  & 90  & 90 & 90\% & 213.8 & 14.2 & $\times$ \\
Gemini Studio & 100  & 80  & 80  & 87\% & 49.3 & 13.0 & 100 & 90  & 70  & 83\% & 108.0 & 15.5 & $\times$ \\
CodeBuddy     & 70  & 60  & 50  & 60\%  & 195.8 & 14.0 & 70  & 60  & 50  & 60\%  & 414 & 11.7 & $\times$ \\
\midrule
\multicolumn{14}{c}{\textit{Ours (Autonomous)}} \\
\midrule
Ours w/o HEG  & 100 & 80  & 70  & 83\%  & 480.0 & 11.2& 100 & 70  & 70  & 80\%  & 465.0 & 16.0 & $\times$ \\
\textbf{Ours (Full)} & \textbf{100} & 80 & 80 & 87\% & 257.0 & 11.2 & \textbf{100} & 70 & 70 & 80\% & 198.0 & 16.2 & \checkmark \\ 
\bottomrule
\end{tabular}
\caption{\textbf{Detailed Metrics: Intermediate Tasks.} Comparison on City Sim (Resource Management) and Snake++ (Event Driven). Note the sharp decline in \textit{Gemini Studio}'s rule adherence due to context drift, while Contract-Coding maintains high fidelity.}
\end{table*}

\begin{table*}[t]
\centering
\label{tab:roguelike_detail}
\resizebox{\textwidth}{!}{%
\begin{tabular}{l|cc|ccc|c|c|c|l}
\toprule
\multirow{2}{*}{\textbf{Method}} & \multicolumn{2}{c|}{\textbf{Arch. Integrity (\%)}} & \multicolumn{3}{c|}{\textbf{Func. Fidelity (\%)}} & \textbf{Overall} & \textbf{Time} & \textbf{Qual.} & \multirow{2}{*}{\textbf{Key Failure Mode}} \\
 & Struc. & Dep. & Exec. & Inter. & Rules & \textbf{Success} & (s) & \textbf{Files} & \\ 
\midrule
\multicolumn{10}{c}{\textit{Legacy Multi-Agent Frameworks}} \\
\midrule
MetaGPT & 60 & 40 & 20 & 10 & 0 & 10\% & 261.0 & 10.9 & Implementation Sparsity \\
ChatDev & 0 & 10 & 20 & 10 & 0 & 10\% & 144.2 & 13.9 & Flat Structure Fallacy \\
FLOW & -- & -- & 0 & 0 & 0 & 0\% & 90.6 & 1.0 & Aggregative Collapse \\
OpenHands & 80 & 50 & 70 & 10 & 10 & 30\% & 310.7 & 14.1 & Lost-in-the-Middle \\
\midrule
\multicolumn{10}{c}{\textit{Commercial AI IDEs}} \\
\midrule
Lingma & 100 & 60 & 40 & 40 & 20 & 33\% & 619.6 & 24.7 & Logical Detachment \\
Trae & 100 & 95 & 80 & 30 & 30 & 47\% & 217.9 & 23.9 & Unreachable Logic \\
Gemini Studio & 80 & 80 & 70 & 70 & 50 & \textbf{63\%} & \textbf{72.1} & 16.6 & Probabilistic Drift \\
CodeBuddy & 100 & 88 & 60 & 40 & 20 & 40\% & 422.2 & 19.6 & Calling Hallucination \\
\midrule
\multicolumn{10}{c}{\textit{Ours (Autonomous)}} \\
\midrule
\textbf{Ours (Full)} & \textbf{100} & 92 & \textbf{90} & 20 & 30 & 47\% & 232.0 & 14.2 & Local Logic Error \\ 
\bottomrule
\end{tabular}%
}
\caption{\textbf{Detailed Results on Roguelike.} High-complexity tasks trigger distinct pathological behaviors in baselines.}
\end{table*}

\section{Appendix: Ablation Experiments}
\label{app:ablation}

\subsection{Contract Auditing}
To validate the Language Contract as an essential structured interface specification, we removed the global contract mechanism (``Ours w/o Contract''). In this setting, agents operate without a Single Source of Truth. 

\paragraph{Impact on Performance.} 
As shown in Table \ref{tab:ablation}, removing the Contract causes a precipitous drop in Functional Success (100\% $\rightarrow$ 65\% in Plane Battle). 

\paragraph{Diagnostic Analysis.} 
To identify the cause, we utilize the Task Existence $E(C)$ metric. Without the Contract to explicitly enumerate and track sub-tasks, the system suffers from severe ``Task Forgetting,'' where the Existence rate drops to 82\%. This confirms that the Contract is critical for preventing task omission in long-horizon generation; without it, parallel agents lose track of requirements, directly leading to system failure.

\begin{table}[h]
\centering
\small
\label{tab:ablation}
\resizebox{\columnwidth}{!}{%
\begin{tabular}{lcccc}
\toprule
\multirow{2}{*}{\textbf{Method}} & \multicolumn{2}{c}{\textbf{Overall Success (\%)}} & \multicolumn{2}{c}{\textbf{Diagnostic}} \\ 
\cmidrule(lr){2-3} \cmidrule(lr){4-5} 
& Plane Battle & Gomoku & Consistency $(V(C))$ & Existence $(E(C))$\\ 
\midrule
Ours w/o Contract & 65 & 85 & - & 82\%(Low) \\
\textbf{Ours (Full)} & \textbf{100} & \textbf{100} & \textbf{100\% (High)} & \textbf{100\% (High)}\\
\bottomrule
\end{tabular}%
}
\caption{\textbf{Component Ablation Study.} Removing the Language Contract significantly impacts Overall Success and Task Existence.}
\end{table}

\subsection{Model Agnosticism Study}
\label{sec:ablation_model}

To verify whether the performance of Contract-Coding stems from the architectural superiority of our paradigm rather than reliance on a specific proprietary model family (i.e., GPT-4o), we conducted a \textbf{Model Agnosticism} ablation. We replaced the underlying LLM with \textbf{Qwen-Plus}, a leading non-OpenAI model, while keeping all hyper-parameters and the Contract-Driven workflow unchanged.

\paragraph{Results.} 
As detailed in Table \ref{tab:model_ablation}, the framework maintained a \textbf{100\% Functional Success Rate} on both tasks when powered by Qwen-Plus. This performance parity is significant: it confirms that the \textbf{Contract-Driven Hierarchical Graph} effectively lowers the reasoning threshold required for repository-scale generation. By decoupling high-entropy architectural planning into low-entropy, auditable interface constraints, our paradigm enables non-GPT-4 models to achieve SOTA performance. This proves that the system's robustness is derived from its \textit{methodological topology}, not merely the latent knowledge of a specific proprietary LLM.

\subsection{Analysis of Parallel Efficiency)}
It is imperative to acknowledge that commercial AI IDEs (e.g., CodeBuddy, Lingma) benefit from significant Infrastructure Advantages, including proprietary low-latency inference engines, which are unavailable to open-source agent frameworks. Despite this physical disparity in raw token generation speed, our method achieves competitive end-to-end efficiency through Architectural Parallelism. As illustrated in Table \ref{tab:main_results}, compared to the heavyweight IDE Trae (393.0s), which suffers from high latency and generates notably bloated code (Files=14.1), our method is over 3x faster (117.7s) while producing significantly cleaner structures (Files=6.2). Furthermore, while optimized tools like CodeBuddy exhibit faster raw execution (90.8s), but they are more inclined to generate simple code(Files=2.9). By transforming the topological structure from a chain to a graph, our Hierarchical Execution Graph (HEG) effectively hides the latency of complex modules, enabling a rigorous, contract-driven multi-agent system to compete with infrastructure-optimized commercial products.

\begin{algorithm}[t]
\caption{Contract-Driven Orchestration \& Auditing} 
\label{alg:auditing}
\begin{algorithmic}[1]
\REQUIRE User Intent $\mathcal{I}$, Max Steps $T$
\STATE $\mathcal{C} \leftarrow \text{SynthesizeContract}(\mathcal{I})$ \COMMENT{Drafting \& Rectification}
\WHILE{$\mathcal{G}$ has unfinished nodes}
    \STATE $\textit{tasks} \leftarrow \text{GetReadyTasks}(\mathcal{C})$ \COMMENT{Topological parallel set}
    \FOR{\textbf{each} $\tau_i \in \textit{tasks}$ \textbf{do in parallel}} 
        \STATE $File_i \leftarrow \text{WorkerAgent}(\tau_i, \mathcal{C})$ \COMMENT{Context is restricted to $\mathcal{C}$}
        \STATE $\delta \leftarrow \text{Auditor}.\text{Compare}(File_i, \mathcal{C}.\text{schema})$
        \IF{$\delta = \emptyset$} 
            \STATE $\mathcal{R}.\text{Commit}(File_i)$; $\text{MarkDone}(\tau_i)$
        \ELSIF{$\delta$ is Critical Mismatch}
            \STATE $\text{Reject}(File_i)$; $\text{Feedback}(\tau_i, \delta)$
        \ELSE
            \STATE $\mathcal{C}.\text{Update}(\delta)$ \COMMENT{Adaptive Contract Patching}
        \ENDIF
    \ENDFOR
\ENDWHILE
\end{algorithmic}
\end{algorithm}

\begin{table}[h]
\centering
\small
\resizebox{\linewidth}{!}{
\begin{tabular}{llccc}
\toprule
\textbf{Task} & \textbf{Backbone Model} & \textbf{Success Rate} & \textbf{Diff.} & \textbf{Outcome} \\
\midrule
\multirow{2}{*}{\textbf{Plane Battle}} & GPT-4o (Main) & 100\% & - & \multirow{2}{*}{\textbf{Parity}} \\
 & Qwen-Plus (Variant) & \textbf{100\%} & 0\% & \\
\midrule
\multirow{2}{*}{\textbf{Gomoku}} & GPT-4o (Main) & 100\% & - & \multirow{2}{*}{\textbf{Parity}} \\
 & Qwen-Plus (Variant) & \textbf{100\%} & 0\% & \\
\bottomrule
\end{tabular}
}
\caption{\textbf{Robustness Across Model Families.} The framework achieves identical stability with Qwen-Plus, demonstrating model agnosticism.}
\label{tab:model_ablation}
\end{table}

\section{Appendix: Forensic Analysis of Failure Cases}
\label{sec:app_failure_analysis}

\begin{figure*}[t] 
   \begin{contractbox}{Listing 1: The Structure of the Language Contract}
\begin{lstlisting}[language=yaml]
# ==========================================
# SECTION 1: Product Requirement Document
# ==========================================
Product Requirement Document:
  1.1 Project Overview: |
    [High-level summary of the project goals and scope...]

  1.2 User Stories:
    - "[User Story 1]"
    - "[User Story 2]"

  1.3 Constraints:
    - "[Technical or Design Constraint 1]"
    - "[Constraint 2]"

# ==========================================
# SECTION 2: Technical Document
# ==========================================
Technical Document:
  2.1 Project Structure:
    - "[Directory/File Structure Tree]"

  2.2 Global Shared Knowledge:
    - "[Shared Constants, Configs, or Global States]"

  2.3 Dependency Relationships:
    - "[Mermaid]"

  2.4 Symbolic Api Specifications:
    - File Path: "[File Path]"
      Owner:     "[Agent Name]"
      Version:   "[Number]"
      Status:    "[Status]"
      
      Classes:
        - Class Name: "[Class Name]"
          
          Attributes:
            - Name: "[Attribute Name]"
              Type: "[Type]"
              Description: "[Description]"
          
          Methods:
            - Signature: "def [Name]([Param Name]: [Type]) -> [Type]"
              Docstring: "[Briefly explain logic and behavior]"
\end{lstlisting}
\end{contractbox}
    \caption{Visualizing the Language Contract. The Contract bridges the Product Requirements Documents (PRD) and Technical Manifold (API Specs). This structure allows the \textit{Contract-Guided Auditing} mechanism to strictly validate implementation compliance via parsing, ensuring parallel context consistency.}
    \label{fig:contract_example}
\end{figure*}

We conducted a manual code inspection of the failed repositories to identify the root causes of the performance gaps reported in RQ4. This section details the specific anti-patterns observed in each baseline.

\subsection{Forensics of Academic Baselines}

\paragraph{MetaGPT: The ''Hollow Skeleton'' Phenomenon.}
Inspection of the repositories generated by MetaGPT revealed a disconnect between the Architect Agent's planning and the Engineer Agent's execution. In approximately 40\% of cases, the directory structure was perfectly instantiated, yet critical logic files (such as \texttt{entities/player.py}) contained only placeholder comments or empty class definitions (e.g., \texttt{class Player: pass}). This suggests that without a persistent constraint like our Language Contract to enforce content density, downstream agents tend to minimize output length to satisfy the immediate prompt.

\paragraph{ChatDev: The Reflection-Action Gap.}
ChatDev exhibited a specific grounding failure related to file system operations. The generated code consistently utilized relative imports assuming a nested directory structure (e.g., \texttt{from core.game import Game}), yet the agent saved all files physically into the root directory. Logs indicate that the Test phase correctly identified the \texttt{ModuleNotFoundError}, but the agent repeatedly attempted to patch the Python import statements rather than performing the necessary OS-level \texttt{mkdir} actions. This highlights a critical limitation in the agent's Action Space.

\paragraph{FLOW: Aggregative Context Collapse.}
FLOW's failure mode was characterized by the inability to maintain file separation. The final aggregation node frequently concatenated logic from multiple parallel sub-tasks into a single, incoherent script. This resulted in massive namespace conflicts and syntax errors, effectively collapsing the intended multi-module architecture into a non-executable monolithic block.

\subsection{Forensics of Commercial AI IDEs}

\paragraph{Gemini AI Studio: Probabilistic Drift.}
Repositories generated by Gemini AI Studio demonstrated the risks of high-speed, open-loop generation. In 20\% of failures, the model hallucinated the user's high-level intent (e.g., generating a generic Grid Puzzle instead of the requested Roguelike). In another 20\%, we observed ``Interface Drift,'' where early-defined modules and late-generated modules used mismatched method signatures. This suggests that without an explicit Contract, the model's internal attention mechanism suffers from decay over long generation sequences.

\paragraph{Lingma: Logical Detachment.}
Lingma's output was characterized by subtle integration bugs that hint at context loss. Specifically, modules frequently lacked necessary error handling, leading to silent runtime crashes. Furthermore, we observed recurring violations of Object-Oriented principles, such as classes failing to implement abstract methods defined in base classes.

\paragraph{Trae: The ``Orphaned Logic'' Paradox.}
While Trae generated the most visually complex code structures, it suffered from ``Orphaned Logic.'' Inspection revealed sophisticated \texttt{StateManager} classes that were fully implemented but never instantiated or invoked within the main \texttt{GameEngine} loop. Similarly, UI rendering loops often failed to poll the game logic state, resulting in a ``frozen'' interactive window despite correct rendering code. This indicates a failure in synthesizing the holistic system flow, where individual components are high-quality but their orchestration is disconnected.

\section{Appendix: Human Evaluation Protocol and Metrics}
To ensure the rigor of our "greenfield" task evaluations, we established a standardized triple-blind human scoring protocol. Each repository was evaluated by three independent reviewers based on the following operational criteria:

\textbf{1. Executability ($E_x$):} 
Measured as a binary pass/fail. A repository is marked as \textit{pass} if it satisfies: (a) Zero \textit{ModuleNotFoundError} after standard pip installs; (b) Successful execution of \textit{main.py} without runtime Traceback for at least 60 seconds of idle time.

\textbf{2. Interactivity ($I_n$):} 
Evaluates the feedback loop between the system and human input. Reviewers followed a fixed action sequence:
\begin{itemize}
    \item \textbf{Gomoku:} Placing a stone on the grid results in a state change and triggers an AI response.
    \item \textbf{Plane Battle:} Directional keys accurately move the sprite; the "Fire" key instantiates bullet objects.
    \item \textbf{City Sim:} Mouse interaction with the UI menu allows selecting building types; clicking the grid successfully places structures and updates the visual HUD.
    \item \textbf{Snake++:} Directional inputs change the snake's heading instantly; activating "Speed Boost" results in a visible acceleration without input lag.
    \item \textbf{Roguelike:} The character moves between rooms and interacts with enemy items (e.g., orc).
\end{itemize}

\textbf{3. Rule Adherence ($R_a$):} 
Checks the logical integrity of game-specific mechanics:
\begin{itemize}
    \item \textbf{Gomoku:} The system correctly identifies 5-in-a-row as a terminal winning state.
    \item \textbf{Plane Battle:} Collision between the player and projectiles triggers health reduction or game-over logic.
    \item \textbf{City Sim:} The resource dependency loop is enforced: Houses generate tax revenue if and only if sufficient Energy exists; Power Plants correctly deplete Money to build.
    \item \textbf{Snake++:} Modular power-up logic functions as defined: "Shields" prevent game-over on wall collision, and "Teleportation Walls" correctly wrap entity coordinates.
    \item \textbf{Roguelike:} Map generation adheres to procedural constraints (e.g., rooms are traversable). 
\end{itemize}

\section{Appendix: The Greenfield-5 Benchmark Specification}
\label{app:benchmark}

We introduce \textbf{Greenfield-5}, a benchmark suite designed to stress-test autonomous agents on "Vibe Coding" scenarios. The suite is open-sourced at \url{https://anonymous.4open.science/r/ContractCoding}.

\subsection{Task Definitions}
Each task is defined by a high-level, ambiguous user prompt (The "Vibe") and a set of functional requirements. 

\begin{itemize}
    \item \textbf{Gomoku (Logic-Intensive):} Write a Gomoku program with AI that allows players to play against AI.
    \item \textbf{Plane Battle (Action-Oriented):} Build a flying battle game where airplanes can move up, down, left, and right, and also fire bullets. Bullets can hit enemies, and enemies will die if hit. Players need to control the airplane to avoid enemy bullets until all enemies are eliminated or the player is hit and killed.
    \item \textbf{City Sim (Resource Management):} Project: Micro City Sim
    Objective: Create a relaxing grid-based city builder in Python/Pygame focused on resource balancing.
    Core Features Needed:
    The Economy Loop: I need a simulation where Money and Energy are calculated in real-time.
    Houses generate Money (Tax) but consume Energy.
    Power Plants generate Energy but cost Money to build.
    If Energy is low, Houses stop paying taxes. (This logic needs to be robust).
    Building System: I want to select different structures from a menu and click the grid to place them.
    Include at least: Roads, Residential Zones, and Industrial Zones.
    Make it easy to add more building types later (Implicitly asks for polymorphism).
    Visual Feedback: A HUD that shows my current resources and alerts me if I'm out of power. The map should look clean.
    Requirements:
    Above 10 files.
    \item \textbf{Snake++ (Event-Driven):} Project: Snake Grand-Master Objective: Build a highly extensible Snake game using Python and Pygame. Functional Requirements:
Advanced Movement: Standard snake growth logic with support for "Speed Boost" and "Teleportation Walls".
Power-up Factory: A modular system to spawn different items (Apple for growth, Magnet for distance, Shield for wall-clip).
Level/Map System: Different arena layouts (e.g., Box, Tunnel, Maze).
Leaderboard \& Save System: A module to handle high-score persistence and rank calculation. Architecture Constraint: The repository must contain 13-15 files.
    \item \textbf{Roguelike (System Architecture):} Project: Abyssal Echoes (Roguelike)
Objective: Generate a repository-level Python project using Pygame. The architecture must be highly modular to support the following complex features:
Procedural Map System: An isolated algorithm module generating rooms and tunnels. Data must be separated from rendering.
Event-Driven UI: A Message Log and HUD that updates via an Event Bus, never by direct coupling to game logic.
Polymorphic Entities: Implement Player, Orc (Melee), and Mage (Ranged/Fleeing AI) sharing a base class.
Interaction System: An inventory system where items can trigger effects across different modules (e.g., a Scroll that reveals the Fog of War map layer).
Constraint: The codebase must be split into at least 15 files. 
\end{itemize}

\subsection{Evaluation Protocol and Metrics}
To ensure rigorous comparison, we employ a dual-layered evaluation harness combining static structural analysis and dynamic functional execution.

\paragraph{1. Static Structural Analysis.} 
Before execution, we verify the repository's topological health to quantify the alleviation of "Architectural Collapse."
\begin{itemize}
    \item \textbf{Architectural Fidelity ($S_{arch} \in [0, 1]$):} Measures the structural alignment with the ground-truth reference architecture. Let $F_{gen}$ and $F_{ref}$ be the sets of file paths in the generated and reference repositories, respectively. We utilize the F1-score of the file set match:
    \begin{equation}
        S_{arch} = \frac{2 \cdot |F_{gen} \cap F_{ref}|}{|F_{gen}| + |F_{ref}|}
    \end{equation}
    This metric penalizes both "Hollow Skeleton" (missing functionality) and "Bloated" (redundant function files) hallucinations.
    
    \item \textbf{Linkage Consistency ($S_{link} \in [0, 1]$):} Evaluates the validity of inter-file symbol resolution. We parse the Abstract Syntax Tree (AST) of all Python files to extract the set of internal import statements $\mathcal{I}$. An import $i \in \mathcal{I}$ (e.g., \texttt{from A import B}) is valid if and only if file \texttt{A} exists and defines symbol \texttt{B}.
    \begin{equation}
        S_{link} = \frac{\sum_{i \in \mathcal{I}} \mathbb{I}(i \text{ is valid})}{|\mathcal{I}|}
    \end{equation}
    High $S_{link}$ indicates a rigorously decoupled and connected dependency graph.
\end{itemize}

\paragraph{2. Dynamic Functional Metrics.}
For repositories that pass basic syntax checks, we calculate the \textbf{Overall Success Score ($S_{overall}$)} as the arithmetic mean of three execution sub-metrics:
\begin{equation}
    S_{overall} = \frac{1}{3} (S_{exec} + S_{inter} + S_{rule}) \times 100\%
\end{equation}
\begin{itemize}
    \item \textbf{Executability ($S_{exec} \in \{0, 1\}$):} The repository must install dependencies (via \texttt{pip}) and launch the entry point without runtime crashes (e.g., \texttt{Traceback}, \texttt{ModuleNotFoundError}) for a 60-second keep-alive window.
    \item \textbf{Interactivity ($S_{inter} \in [0, 1]$):} Measured via a standardized UI action sequence (e.g., "Click Start" $\to$ "Move Player"). The system scores points for valid state transitions/responses to automated user inputs.
    \item \textbf{Rule Adherence ($S_{rule} \in [0, 1]$):} Deep verification of game-specific logic (e.g., "Health decreases exactly by 10 upon collision"). This detects subtle "Zombie Code" that runs but fails to implement intent.
\end{itemize}

\begin{figure*}[t] 
    \centering
    
    \begin{promptbox}[width=\linewidth]{System Prompt}
\begin{verbatim}
You are an expert agent within a larger, collaborative multi-agent system. Your primary goal is to 
contribute to the overall task by performing your specific role and then passing control to the next 
appropriate agent(s).

# Important Instruct Reminders

1. Do what has been asked; nothing more, nothing less.
2. NEVER create files unless they're absolutely necessary for achieving your goal. This means NO 
documentation (like README.md), configuration, or test files unless you are explicitly told to create 
them.
3. ALWAYS prefer editing an existing file to creating a new one. 
4. Use OpenAI function calling to execute tools.

# Collaboration Guideline

1. **Collaboration is Key**: All agents work together to achieve the project's goals.
2. **Document Management**: You have access to a shared "Collaborative Document". All agents in this 
workflow can read and write to. 
    It is the central place for sharing knowledge, plans, API definitions, file contents, or any other 
    IMPORTANT information.
3. **Document Conciseness**: The content of Collaborative Document should be as concise as possible, 
providing key information or API interfaces.
4. **Context Management**: Keep only necessary information in the document. Remove outdated specifica-
tions, redundant details, and verbose descriptions.
5. **API Minimalism**: API descriptions should include only: endpoint path, method, required parameters
, and response format. Omit lengthy explanations.

## Document Structure
The Collaborative Document MUST contain:

1) Requirements Document (Problem-space, contract-agnostic)

2) Technical Document (Solution-space, file-led)
    - Sub-Tasks(File-based)

# DOCUMENT ACTION LANGUAGE GUIDELINE
The `<document_action>` tag contains a JSON array of action objects. All agents share the SAME 
`Collaborative Document`. `content` is a MARKDOWN string representing the FULL document.
**`update|add`**: Updates or Adds content to the Collaborative Document.
- Section-patch mode (preferred): `content` is a JSON object where keys are section names and values are
MARKDOWN section bodies (NO section heading lines).
  - Allowed section keys:
    - "Project Overview"
    - "User Stories (Features)"
    - "Constraints"
    - "Directory Structure"
    - "Global Shared Knowledge"
    - "Dependency Relationships"
    - "Symbolic API Specifications"
  - The system will apply your section patches onto the current document and keep the overall document in 
  the required MD template format.
  - IMPORTANT: A section patch is a FULL REPLACEMENT of that section's body. If you change only part of a 
  section, you MUST still output the entire final section body, including unchanged lines.
  - WARNING: Partial section patches (e.g., only "* **Status:** DONE" without any "**File:** ...") may 
  be rejected to prevent clobbering the section.
- Example: `[{"type": "update|add", "section": "key", "content": "...markdown to append..."}]`

# INSTRUCTIONS: Your response MUST follow this structure EXACTLY, this is VERY IMPORTANT.
1.  **Thinking Process**: In a `<thinking>` block, provide a step-by-step analysis of the current situ-
ation, your reasoning, and your plan.
2.  **Output**: In an `<output>` block, provide your primary output. A human-readable text summary of 
your work, analysis, or conclusion.
3.  **Document Actions (Optional)**: If you need to modify the shared document, provide a 
`<document_action>` block containing a valid JSON array of action objects. If you don't need to modify 
the document, omit this entire block.
\end{verbatim}
    \end{promptbox}
    
    \label{fig:system_prompt}
\end{figure*}
\begin{figure*}[t] 
    \centering
    
    \begin{promptbox}[width=\linewidth]{Contract Prompt}
\begin{verbatim}
You are the Project Manager responsible for producing a thorough, correct, and task‑driven implementa-
tion plan. Your plan must be contract‑first, dynamically structured (no pre‑committed stack), parallel-
izable, and observable. It must apply across different programming tasks and domains.

### Role Principles
- Dynamic structure: choose only the minimal architecture and tools needed for the current task; 
don’t pre‑commit to a fixed stack.
- Prioritize correctness: Ensure the rationality of task coordination after your decomposition
- Contract‑first: define interfaces and algorithm contracts (paths/methods/data shapes/types).
- Parallelization & simplicity: decompose into concurrent modules while minimizing complexity.
- Focus only on essential implementation tasks: Each task should produce concrete deliverables.
- Read the Collaborative Document and the user task. Produce a plan that is implementable, explicitly 
scoped, and adaptable.
- Include only what is necessary for the current task; every choice (tech/structure) must be justified 
by necessity.

###Task decomposition
- **Keep the workflow streamlined.**
- **Maintain clarity and logical flow.** Decomposition should be intuitive, avoiding redundant or tedi-
ous steps.
- **Prioritize core functions.** Focus on the main features required, rather than edge situations or ad-
vanced features.
- **Decomposition principle.** The decomposition of subtasks can be based on the files and their imple-
mented functions.

### CRITICAL INSTRUCTION: Document Creation
- You MUST use the `document_action` tool with type `add` to CREATE the initial Collaborative Document.
- The content of the document MUST be placed INSIDE the JSON list in <document_action>.
- The document MUST follow the template below EXACTLY.

**Collaboration Document Template:**
{Contract Template}
\end{verbatim}
    \end{promptbox}
    
    \label{fig:project_prompt}
\end{figure*}

\end{document}